\documentclass[a4paper,twoside,notoc,11pt]{JHEP3}

\usepackage{epsfig,multicol}
\usepackage{delarray,amsmath,bbm}

\newcommand\fverb{\setbox\pippobox=\hbox\bgroup\verb}
\newcommand\fverbdo{\egroup\medskip\noindent%
                              \fbox{\unhbox\pippobox}\ }
\newcommand\fverbit{\egroup\item[\fbox{\unhbox\pippobox}]}
\newbox\pippobox

\newcommand{\nn}{\nonumber}
\newcommand{\beq} {\begin{equation}}
\newcommand{\eeq} {\end{equation}}
\newcommand{\beqa} {\begin{eqnarray}}
\newcommand{\eeqa} {\end{eqnarray}}

\newcommand{\ie}{{\it i.e.}}
\newcommand{\eg}{{\it e.g.}}
\newcommand{\cf}{{\it cf.}}

\newcommand{\ieps}{i\varepsilon}
\newcommand{\order}[1]{${\cal O}\left(#1 \right)$}
\newcommand{\morder}[1]{{\cal O}\left(#1 \right)}
\newcommand{\eq}[1]{(\ref{#1})}

\newcommand{\ave}[1]{\langle{#1}\rangle}
\newcommand{\com}[2]{\left[{#1},{#2}\right]}
\newcommand{\tr}{\mathrm{Tr}\,}

\newcommand{\mati}[2]{{_{#1}^{\ {#2}}\,}}

\newcommand{\pvec}{{\bf p}}

\newcommand{\qu}{{\rm q}}

\newcommand{\qb}{{\rm\bar q}}

\newcommand{\qq}{\qu\qb\ }
\newcommand{\chisb}{{\raisebox{0.8mm}{\hbox{$\chi$}}${\textstyle SB}$}}

\newcommand{\halft}{{\textstyle \frac{1}{2}}}

\newcommand{\Lslash}[1]{ \parbox[b]{1em}{$#1$} \hspace{-0.8em}
                                  \parbox[b]{0.8em}{
\raisebox{0.2ex}{$/$}}}
\newcommand{\Slash}[1]{ \parbox[b]{0.6em}{$#1$} \hspace{-0.55em}
                                  \parbox[b]{0.55em}{
\raisebox{-0.2ex}{$/$}}}

\title{\center{QCD Green functions in a gluon field}}

\author{Paul Hoyer\thanks{Research supported in part by the
Academy of Finland through grant 102046.}\\
              Department of Physical Sciences and Helsinki Institute of
              Physics\\
              POB 64, FIN-00014 University of Helsinki, Finland \\
              E-mail: \email{paul.hoyer@helsinki.fi}}
\author{St\'ephane Peign\'e\\
              Laboratoire d'Annecy-le-Vieux de Physique Th\'eorique\\
              LAPTH, CNRS, UMR 5108, Universit\'e de Savoie\\
              B.P. 110, F-74941 Annecy-le-Vieux Cedex, France\\
              E-mail: \email{peigne@lapp.in2p3.fr}}
\preprint{HIP-2004-53/TH \\ LAPTH-1073/04 \\  \hepph{0410235}
}

\abstract{
We formulate a dressed perturbative expansion of QCD, where the
standard diagrams are evaluated in the presence of
a constant external gluon field whose magnitude is gaussian distributed. 
The approach is Poincar{\'e} and gauge
invariant, and modifies the usual results for hard processes
only by power suppressed contributions. Long distance
propagation of quarks and gluons turns out to be inhibited due to
a branch point singularity instead of a pole at $p^2=0$ in the
quark and gluon propagators.
The dressing keeps the (massless) quarks in $\qq$ fluctuations of
the photon at a finite distance from each other.
}


\begin{document}

\section{Introduction}

The presence of a gluon and quark `condensate' in the QCD ground state
is a plausible reason for the observed long distance properties of
QCD \cite{svz, ds}. The condensate apparently prevents
quarks and gluons from propagating
over long distances, while acting as a superfluid for color singlet
hadrons. In this work we model the gluon condensate effects by
coupling quarks and gluons to a `vacuum' gluon field $\Phi$ which
is taken as a constant in space-time in a covariant gauge.
Translation invariance is thus automatically satisfied.
We integrate over the Lorentz and color components of
$\Phi_\mu^a$ with a gaussian weight. This ensures Lorentz and gauge
invariance and introduces a dimensionful parameter $\Lambda$ which
characterizes the magnitude of the vacuum field. In effect, we
modify QCD perturbation theory (PQCD) by expanding around
non-vanishing gluon field configurations.

In the limit of a large number of colors $N \to \infty$ with $g^2N$
fixed \cite{thooft}, we are able to find the exact expressions for the  
gluon
and (massless) quark propagators in a `dressed tree' approximation. The
vacuum field effects are resummed to all orders, whereas
perturbative loop corrections are neglected. The dressed tree
propagators have a $1/\sqrt{p^2}$ branch cut
instead of a pole at
$p^2=0$, and consequently decay with time $t$ as $1/\sqrt{t}$.
For $|p^2| \gg \mu^2 = g^2N\Lambda^2$ the dressed
propagators approach the free ones. Hence the short distance structure
of PQCD is unaffected by the vacuum field.
It is gratifying that the momentum dependence of
our dressed quark and gluon propagators in the large $N$ limit
turns out to agree qualitatively with the results of lattice  
calculations
at $N=3$.

The finite propagation length of the dressed partons appears to  
regularize
the infrared (IR) singularities of the standard PQCD expansion. We
study the dressed (massless) quark loop contribution to the photon
self-energy. Zero-momentum gluons can couple to the quark loop even
for spacelike photons due to the appearance of infrared singularities
in Feynman diagrams involving the coupling of at least $4$ external
fields. While the contribution of a specific number of external fields
is thus ill-defined, the loop integral is IR regular when the fully
dressed quark propagator and quark-photon vertex
are used. The effective lower cut-off for the
loop momentum $k$ is  $k^2 \sim \mu^2 = g^2N\Lambda^2$ (in euclidean
space). Thus the dressing indeed `confines' the color singlet
quark pair within a distance $\sim 1/\mu$.
At high photon virtualities $p^2$ the dressing correction is
$\propto 1/p^4$ as in the QCD sum rule framework \cite{svz}, where the
normalization would be given by the vacuum expectation value 
$\ave{\alpha_s F^a_{\mu\nu}F_a^{\mu\nu}}$.

Since the elementary constituents of QCD are confined
their Green functions need not have a standard analytic structure.
Our dressed quark and gluon propagators indeed have branch cuts in  
$p^2$.
The present framework may thus give insights into how an S-matrix
consistent with general principles can be constructed in a confining  
theory.
A first step in this direction will be to identify the
asymptotic states of our framework.

This work represents a further development of our previous work  
\cite{hp},
where some of the results on the dressed quark propagator and photon
self-energy were already presented.

\section{Coupling quarks and gluons to the vacuum field}
\label{lagsec}

Gauge invariant couplings of quarks and gluons to the
vacuum field $\Phi$ can be found by shifting the gluon field,
\beq
\label{CV}
A^\mu \to A^\mu + \Phi^\mu
\eeq
The quark part $\bar\psi i\Lslash{D}\psi$ of the massless QCD
lagrangian (with $D_{\mu} = \partial_{\mu}  + i g A_{\mu}$ the
covariant derivative) then generates the coupling
\beq
\label{quarkphi}
{\cal L}_{\Phi {\rm q}} = -g \bar\psi \Phi^{\mu} \gamma_{\mu} \psi
\eeq
which we shall use. It is invariant under the gauge transformation
$\psi \to U(x) \psi$ with $U(x) \in {\rm SU}(N)$ provided $\Phi$
transforms as
\beq
\label{phitrans}
\Phi^\mu \to U(x) \Phi^\mu U(x)^\dag
\eeq
This transformation law is consistent with that of the shifted gluon
field,
\beq
A^\mu + \Phi^\mu \to U (A^\mu  + \Phi^\mu) U^\dag
+ \frac{i}{g} \left(\partial^\mu U\right)U^\dag
\eeq
Under the shift \eq{CV} the field strength tensor
\beq
\label{fmunu}
F_{\mu\nu} = \partial_\mu A_\nu -\partial_\nu A_\mu
+ig\com{A_\mu}{A_\nu}
\eeq
becomes
\beqa
F_{\mu\nu} &\to&  F_{\mu\nu} +  F^{\Phi}_{\mu\nu}
+ \Phi_{\mu\nu}
\label{fmunushift} \\
F^{\Phi}_{\mu\nu} &=& \partial_\mu \Phi_\nu -\partial_\nu \Phi_\mu
+ig( \com{\Phi_\mu}{A_\nu} - \com{\Phi_\nu}{A_\mu}) \\
\Phi_{\mu\nu} &=&  ig \com{\Phi_\mu}{\Phi_\nu}
\eeqa
Since $F_{\mu\nu} \to U F_{\mu\nu} U^\dag$ under a gauge
transformation, each term in the r.h.s. of \eq{fmunushift}
of a given degree in $\Phi^\mu$ transforms similarly:
\beq
F_{\mu\nu} \to U F_{\mu\nu} U^\dag \ \ ;\ \
F^{\Phi}_{\mu\nu} \to U F^{\Phi}_{\mu\nu} U^\dag \ \ ;\ \
\Phi_{\mu\nu} \to U \Phi_{\mu\nu} U^\dag
\label{trans}
\eeq

We wish to study the influence of a low-momentum `vacuum' gluon field
$\Phi$ on quark and gluon propagation without delving into the more
difficult question of how the condensate field is generated. We
therefore ignore the self-interactions of $\Phi$ and consider only the
gauge invariant coupling of the $\Phi$ field to gluons
which is linear in $\Phi$,
\beq
\label{gluonphi}
{\cal L}_{\Phi {\rm g}} = - \tr \left[F^{\mu\nu}  F_{\mu\nu}^\Phi
\right]
\eeq

We shall thus work with the modified (massless) QCD lagrangian,
\beq
\label{lag}
{\cal L}  =  \bar\psi i\Lslash{D} \psi -\frac{1}{2}\tr
\left[F_{\mu\nu}F^{\mu\nu} \right]
- \frac{1}{\xi} \tr \left[ (\partial_{\mu} A^\mu)^2  \right]
- {\bar c}\, \partial^{\mu} D_{\mu} c - \lambda \, g
\bar\psi \Lslash{\Phi} \psi
- \tr \left[F^{\mu\nu}  F_{\mu\nu}^\Phi \right]
\eeq
which includes covariant
gauge fixing and ghost terms and is BRST invariant. Since the quark and
gluon
couplings to $\Phi$ are separately gauge invariant there
is no constraint on their relative weight $\lambda$. For simplicity
we choose $\lambda = 1$ in the following.

As the field $\Phi$
is meant to describe the long wavelength (vacuum) effects we
take it to carry zero momentum, \ie, to be independent of the
coordinate $x$ (in the covariant gauge specified by \eq{lag}).
Translation invariance is thus guaranteed. In order 
to preserve Lorentz (and gauge) invariance we average over all Lorentz
(and color) components of $\Phi_\mu^a$ with a gaussian
weight\footnote{The integrals over the time components 
$\Phi_0^a$ are defined by analytic continuation.},
\beq
\label{phiave}
\left( \prod_{}\,\int_{-\infty}^\infty  d\Phi_\mu^a \right)
\exp\left[\frac{1}{2\Lambda^2}
\Phi_\nu^b  \Phi^\nu_b \right]
\eeq
where $\Lambda$ is a parameter with the dimension of mass.
In a perturbative expansion we may interpret \eq{phiave} as giving a
$\Phi$
`propagator'
\beq
\label{phiprop}
i D_{\Phi,\mu\nu}^{ab}(p)  = - \Lambda^2  g_{\mu\nu} \delta^{ab}
(2\pi)^4 \delta^4(p)
\eeq

In the next section we derive explicit expressions for the quark
and gluon propagators in a `dressed tree' approximation, which takes
all interactions with the $\Phi$ field into account, but neglects
perturbative quark and gluon loops.

\section{Dressed quark and gluon propagators in the large $N$ limit}

We now calculate the effects of the interaction terms \eq{quarkphi}
and \eq{gluonphi}
of the zero-momentum vacuum field $\Phi$ on the
quark and gluon propagators. We simplify the topology of the
contributing diagrams by taking the limit of a large number of colors,
$N\to\infty$ with $g^2N$ fixed \cite{thooft}. In addition to the
coupling $g^2N$ we have a parameter $\mu$ with the dimension of
mass,
\beq
\mu^2 = g^2 N \, \Lambda^2
\label{mu}
\eeq
With the lagrangian \eq{lag} the full perturbative
expansion of any quark and gluon Green function $G$ 
is a double sum of the form
\beq \label{pertexp}
G =\sum_{\ell=0}^\infty (g^2N)^\ell \sum_{k=0}^\infty C_{\ell,k}\
\mu^{2 k}
\eeq
Here $\ell$ counts the number of perturbative loops and $2k$ the
number of $\Phi$ couplings \eq{quarkphi} or \eq{gluonphi}. We
evaluate the complete sum over $k$ for $\ell=0$. This is possible
since the $\Phi$ propagator \eq{phiprop} carries zero momentum so no
loop integrals are involved. The remaining sum over $\ell$
involves higher powers of the coupling $g^2N$ and
an increasing number of non-trivial loop
integrals\footnote{We do not consider quark and gluon loops which
are connected to the rest of the diagram only via $\Phi$ lines.
These would generate self-couplings of the $\Phi$ field.}.

\subsection{Dressed quark propagator}

\begin{figure}[b]
\begin{center}
\epsfig{figure=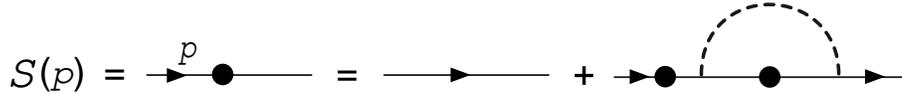,angle=0,width=12cm}
\end{center}
\caption{Implicit equation for the
dressed quark propagator $S(p)$. The dashed line denotes the
$\Phi$ `propagator' (2.13).
\label{quarkDS}}
\end{figure}

Only planar $\Phi$ `loop' corrections to the free quark propagator
contribute in
the large $N$ limit (with $\ell = 0$ in \eq{pertexp}). From the
structure of a general diagram it is readily seen that the dressed quark
propagator $S(p)$ satisfies the Dyson-Schwinger (DS) type equation
shown in
Fig.~\ref{quarkDS}, which reads
\beqa
iS(p) &=& \frac{i}{\Slash{p}}+ C_F(-ig)^2(-\Lambda^2)
\frac{i}{\Slash{p}} \gamma^{\mu} iS(p) \gamma_{\mu} iS(p) \nn \\
\Rightarrow\ \ &&
\Slash{p} S(p) = 1 - \halft \mu^2 \, \gamma^{\mu} S(p) \gamma_{\mu}
S(p)
\label{S0}
\eeqa
where $\mu$ is defined in \eq{mu} and we used $C_F = N/2$ at leading
order in $N$.
The DS equation \eq{S0} generates all one-particle irreducible as well
as reducible planar diagrams.

Lorentz invariance constrains the quark propagator to be of the form 
\beq \label{genform}
S(p) = a_p\,\Slash{p}+b_p
\eeq
where $a_p$ and $b_p$ are functions of $p^2$.
Substituting this in \eq{S0} we get
\beq
\label{condS}
b(1+\mu^2 a) = 0  \ \ {\rm and} \ \ ap^2(1-\mu^2 a) = 1-2\mu^2 b^2
\eeq
A chiral symmetry conserving quark propagator $S_1(p)$
must have $b=0$. The second order equation for $a$ then gives
\beqa \label{aexpr}
a_p &=&
\frac{1}{2\mu^2}\left(1-\sqrt{1-\frac{4\mu^2}{p^2}}\right) \\
\nonumber\\
S_1(p) &=& \frac{2 \Slash{p}}{p^2 + \sqrt{p^2(p^2-4\mu^2)}}
\label{csbcons}
\eeqa
where we chose the sign of the square root so as to ensure that the
propagator approaches the free one in the $p^2 \to\infty$ limit:
\beq \label{asp2bis}
S_1(p)=\frac{1}{\Slash{p}}\left[1+\morder{\frac{\mu^2}{p^2}}\right]\ \ \
{\rm for}\ \ p^2 \to\infty
\eeq
The $p^2 =0$ pole of the
free quark propagator $\Slash{p}/p^2$ has been removed by the dressing.
The dressed propagator $S_1(p)$ given in \eq{csbcons} has
instead branch point singularities at $p^2 =0$ and
$p^2=4\mu^2$. In appendix A (see \eq{Sasymp}) we show that
this removes the quark from the set of asymptotic states,
in the sense that the Fourier transformed propagator vanishes at
large times,
\beq \label{ast}
|S_1(t, \pvec)| \ \ \mathop{\sim}_{|t| \to \infty} \ \
\morder{1/\sqrt{|t|}}
\eeq

Allowing chiral symmetry breaking ({\chisb}), \ie, $b\neq 0$ in
\eq{condS} we find
\beq \label{csbbreak}
S_2(p) = -\frac{1}{\mu^2}\left(\Slash{p} \pm \sqrt{p^2 +
      \mu^2/2}\right) = \frac{1}{2(\Slash{p} \mp \sqrt{p^2 + \mu^2/2})}
\eeq
This solution is singular for $\mu^2 \to 0$ and hence does not have
a power expansion in $\mu^2$ of the form \eq{pertexp}. It emerges
as a `non-perturbative' solution of the implicit equation \eq{S0}. Like
the chiral symmetry conserving solution $S_1$, the propagator
$S_{2}$ has no quark pole, only a branch point at $p^2 = -\mu^2/2$,
where it coincides with $S_1$. Since the solution $S_{2}(p)$ does not 
approach the perturbative propagator
$1/\Slash{p}\ $ at large $p^2$ it must be discarded {\it at short distance.}

In Euclidean space we can estimate the chiral symmetry breaking effect
of using the solution $S_2(p_E)$ for $0 \leq p_E^2 = -p^2 \leq \mu^2/2$
and the $S_1$ propagator for $p_E^2 > \mu^2/2$. The value of the quark
condensate obtained in this way (using the upper sign in \eq{csbbreak}) is
\beq
\ave{\bar \psi \psi} = \int \frac{d^4p_E}{(2\pi)^4} \,
\tr\left[S_2(p_E) \right] \Theta(\mu^2/2 - p_E^2) = - \frac{\mu^3}{60 \,
    \pi^2 \sqrt{2}}
\label{cond}
\eeq
This estimate, ${\ave{\bar \psi \psi}}^{1/3} \simeq -\mu/9.4$,
is an order of magnitude below the generic scale $\mu$.

\begin{figure}[t]
\begin{center}
\epsfig{figure=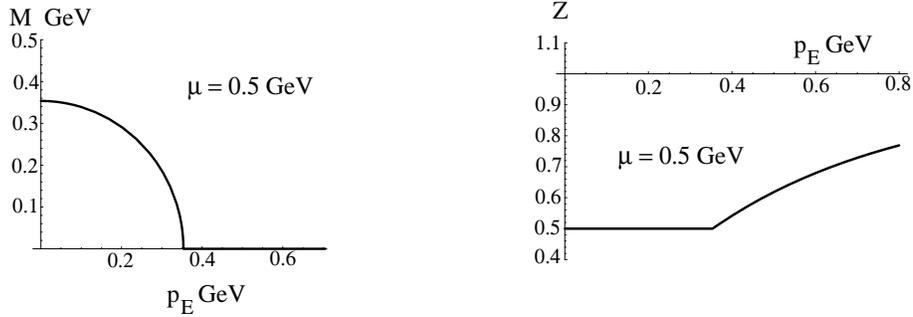,angle=0,width=12cm}
\end{center}
\caption{Numerical results for the quark propagator in the euclidean
region $p_E^2 \equiv -p^2 > 0$. The quantities $M$ and $Z$ are defined
in (3.12).}
\label{figquarkh}
\end{figure}

Expressing the quark propagator as
\beq
S(p) = \frac{Z(p^2)}{\Slash{p} - M(p^2)}
\label{param}
\eeq
we plot in Fig.~\ref{figquarkh} the functions $M(p^2)$ 
and $Z(p^2)$, using $\mu = 500\,\rm{MeV}$ and
the above combination of $S_1$ and $S_2$
in the euclidean $p^2 < 0$ region. Their shapes
are similar to the results obtained in lattice calculations, see
for instance Refs.~\cite{ds} or Fig.~3 of Ref.~\cite{Bonnet-quark}.

Let us mention that 
the above solutions $S_1$ and $S_2$ for the quark propagator
were previously obtained in a different framework \cite{Munczek:1983dx},
and recently within a model assuming the presence of an
$\ave{A_{\mu}^a A^{\mu}_a}$ condensate \cite{li}.
In the rest of this paper we shall use $S(p) = S_1(p)$ 
for all values of $p^2$.

\subsection{Dressed gluon propagator \label{gluesec}}

It is convenient to use the double line color notation \cite{thooft}
for the fields $A^{\mu}$ and $\Phi^{\mu}$ in the limit $N \to
\infty$. The doubly indexed fields are defined in terms of the color
generator matrices $T^a$ of the fundamental representation as
\beq
\label{gluonf1}
{A^\mu}\mati{i}{j} = A^{\mu a} T^a\mati{i}{j}
\eeq
In the covariant gauge defined by \eq{lag} the free propagators and
their graphical representations are then
\beqa
iD_0^{\mu\nu}(p)\mati{i}{j}\mati{k}{l} &=& \ \ \
\mbox{\raisebox{ 
-0.4cm}{\hbox{\epsfxsize=3truecm\epsfbox{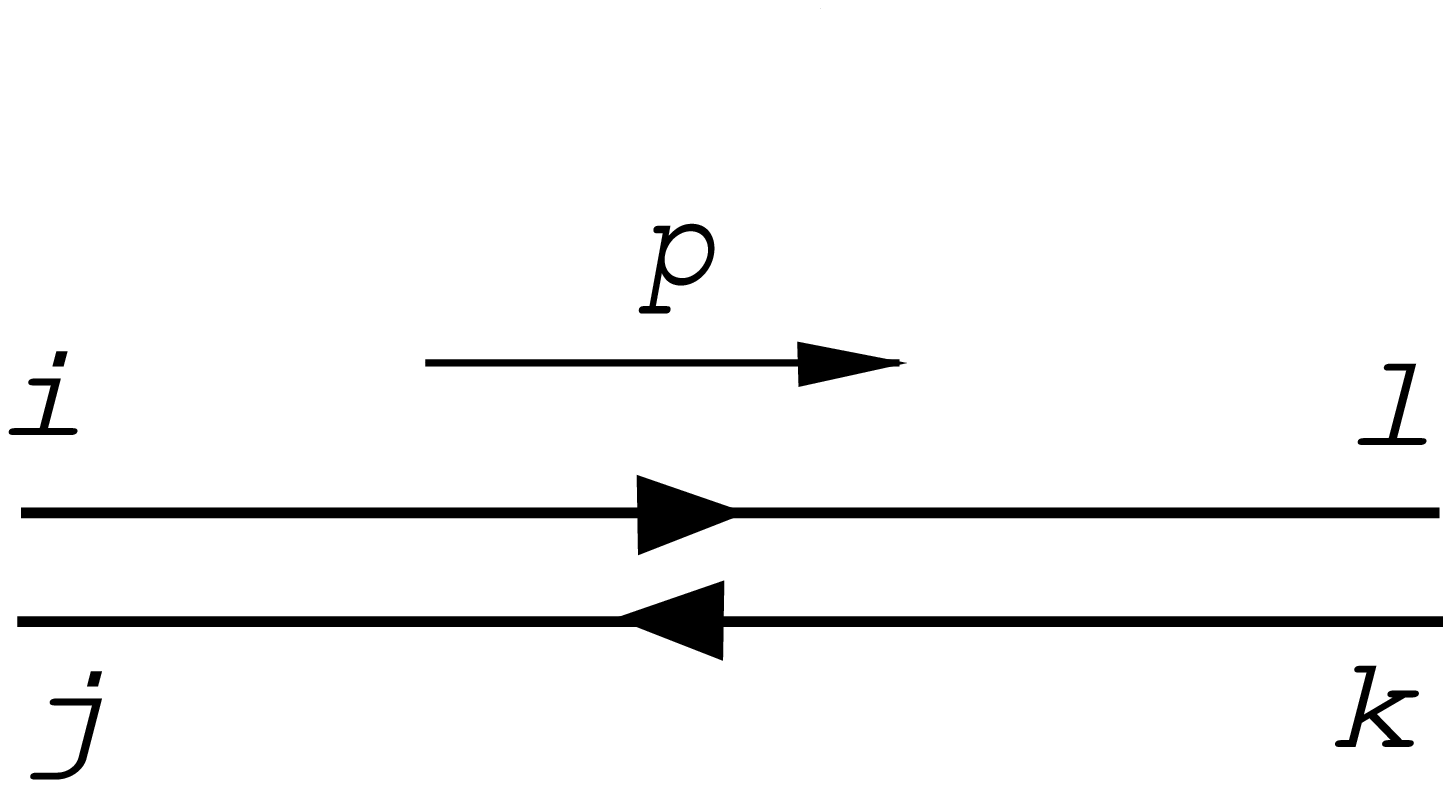}}}}
\ \ \ = \frac{1}{2}\delta\mati{i}{l}
\delta\mati{k}{j}\, \frac{-i}{p^2}
\left[g^{\mu\nu}-(1-\xi)\frac{p^\mu p^\nu}{p^2}\right]
\label{gluonprop} \\
iD_{\Phi}^{\mu\nu}(p)\mati{i}{j}\mati{k}{l} &=& \ \ \
\mbox{\raisebox{-0.4cm}{\hbox{\epsfxsize=3truecm\epsfbox{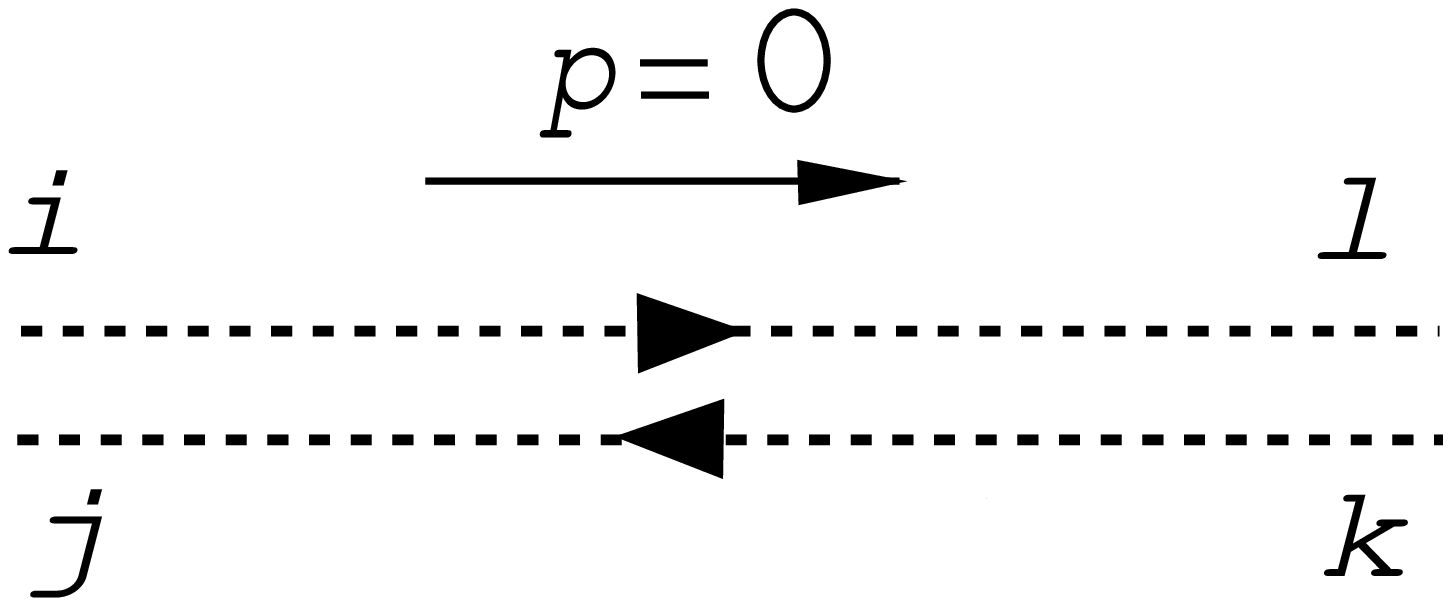}}}}
\ \ \ = -\frac{1}{2}\delta\mati{i}{l} \delta\mati{k}{j}\, \Lambda^2
g^{\mu\nu} \, (2\pi)^4 \delta^4(p)
\eeqa

The interaction term \eq{gluonphi} couples $\Phi$ to two and three
gluons, via $\Phi AA$ and $\Phi AAA$ vertices, respectively.
The $\Phi AAA$
vertex does not, however, contribute in the  dressed tree
approximation (the $\ell=0$ term in \eq{pertexp}), since it
implies at least one perturbative loop integral. The $\Phi AA$ vertex is
\beqa
\label{phiaa}
{\cal L}_{\Phi AA} &=&
-2ig \tr \big( \com{\Phi_\mu}{A_\nu} (\partial^{\mu}A^{\nu}-
    \partial^{\nu}A^{\mu}) \big)  \nn \\
&=& -2ig \tr \big( \Phi_\mu \com{A_\nu}{\partial^{\mu}A^{\nu}}
- \Phi_\mu \com{A^\mu}{\partial^{\nu}A_{\nu}}  \big)
\eeqa
where we used $\partial_\mu\Phi_\nu = 0$ and dropped a total derivative
to obtain the second line. The second term of
\eq{phiaa} vanishes in Landau gauge,
$\partial^{\nu}A_{\nu} =0$. However, we keep both terms in order to
see the dependence on the gauge parameter $\xi$.
In the double line notation the corresponding Feynman rules are:

\beqa
\mbox{\raisebox{-0.6cm}{\hbox{\epsfxsize=4truecm\epsfbox{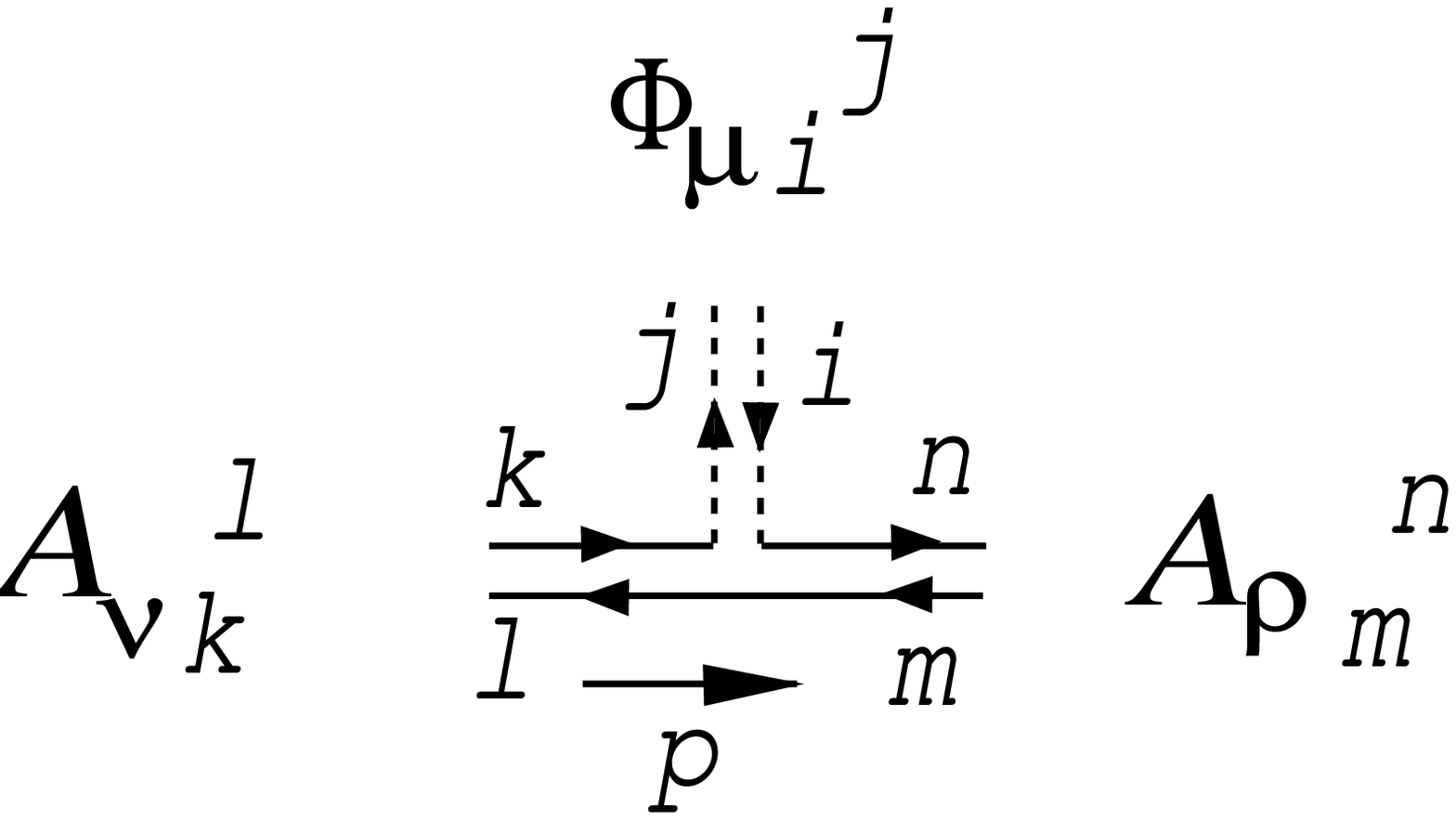}}}}
\hskip .5cm
&=& \ -4ig \,  p_{\mu}\, g_{\nu \rho}\,
\delta\mati{j}{k} \delta\mati{l}{m} \delta\mati{n}{i}
\label{vertex1} \\
\mbox{\raisebox{-0.6cm}{\hbox{\epsfxsize=4truecm\epsfbox{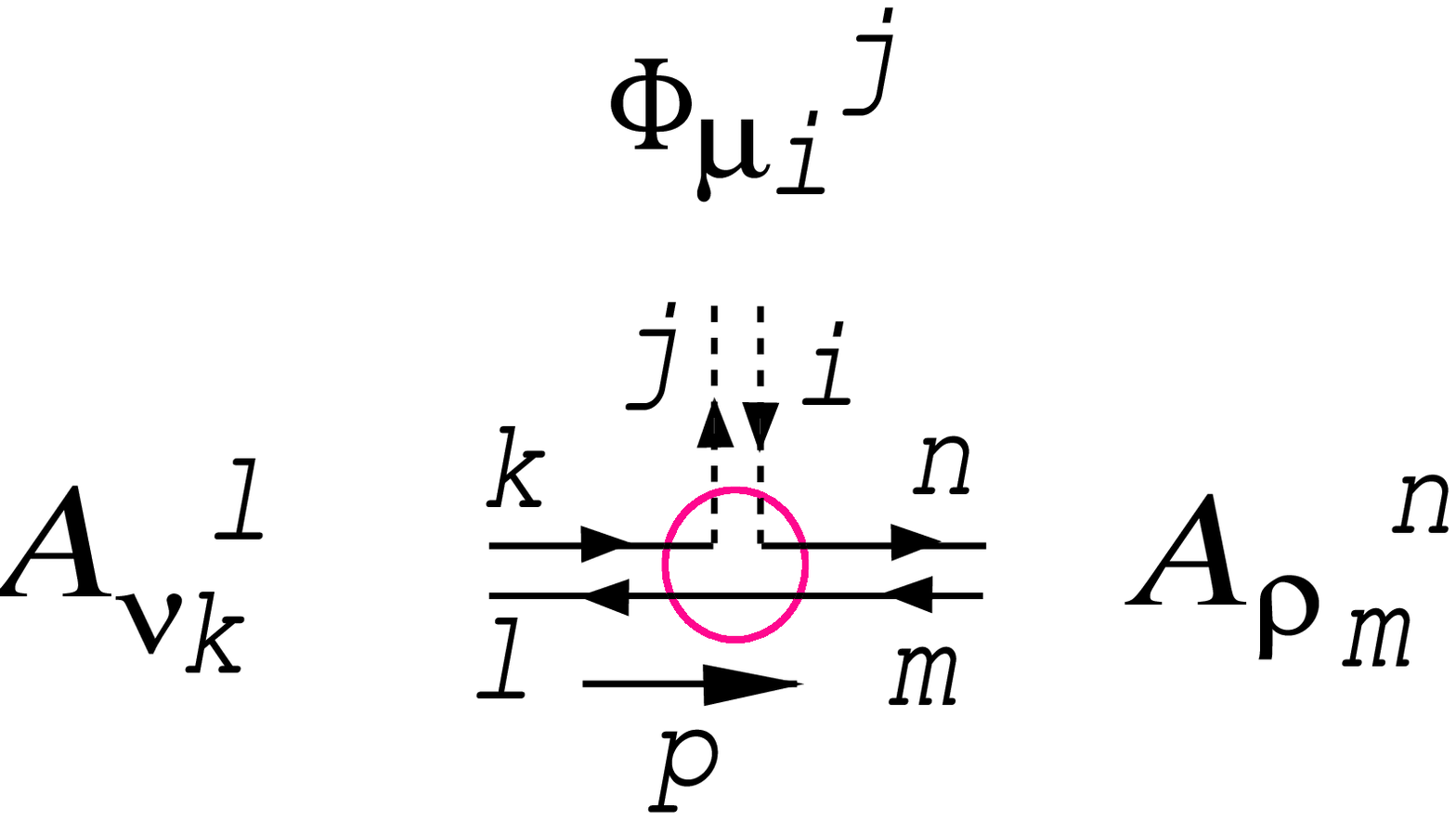}}}}
\hskip .5cm
&=& \ +4ig\, p_{\rho}\, g_{\mu \nu} \,
\delta\mati{j}{k} \delta\mati{l}{m} \delta\mati{n}{i}
\label{vertex3}
\eeqa

\begin{figure}[t]
\begin{center}
\epsfig{figure=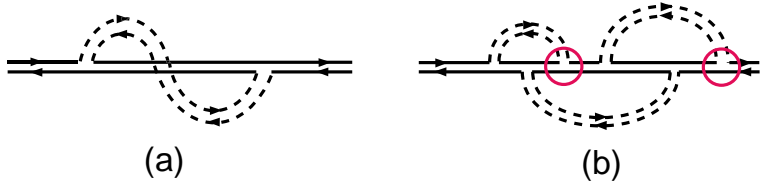,angle=0,width=10cm}
\end{center}
\caption{Diagrams contributing to the gluon propagator
in the double line color notation: a non-planar subleading
contribution (a) and a generic dominant contribution (b) in the large  
$N$
limit.}
\label{doubleline}
\end{figure}

The rules for vertices where the $\Phi$ field attaches to the lower
gluon color line
are the same (up to an overall sign). Since nonplanar diagrams like
the one shown in
Fig.~\ref{doubleline}a do not contribute at large $N$, the $\Phi$ lines
never cross the gluon propagator. A generic diagram contributing to our
dressed tree
approximation is shown in Fig.~\ref{doubleline}b.

One might expect that the dressed tree gluon propagator would satisfy
a DS equation (or a finite set of equations) analogous to the one 
satisfied by the dressed quark propagator (Fig.~1). However, because planar
corrections to the gluon propagator can appear on both sides of the
gluon line the dressed $\Phi gg$ vertex is not proportional to the
bare vertex, as is the case for the quark. This apparently
precludes a finite set of closed DS equations for the gluon propagator
(which in our approach would imply an algebraic solution). By explicitly 
summing all diagrams we indeed find the non-algebraic expression 
\eq{dressedgluon} below.

We first do the calculation in Landau gauge
($\xi =0$), and then evaluate the $\xi$ dependence. In Landau gauge the
free
gluon propagator is transverse,
\beq
\label{gluonlandau}
iD_0^{\mu\nu}(p)\mati{i}{j}\mati{k}{l} = \frac{1}{2}\delta\mati{i}{l}
\delta\mati{k}{j}\, \frac{-i}{p^2} \, P_T^{\mu\nu}(p) \ \ ;\ \
P_T^{\mu\nu}(p) = g^{\mu\nu}-\frac{p^{\mu}p^{\nu}}{p^2}
\eeq
and thus the circled vertex \eq{vertex3} does not
contribute. Independently of the topology of a diagram,
adding one $\Phi$ (double) line always yields the same factor, namely,
\beq
N (-4ig)^2 p^2 \left( \frac{-\Lambda^2}{2} \right)
\left( \frac{-i}{2p^2} \right)^2  = - \frac{2 \mu^2}{p^2}  \equiv x
\eeq
Let $n(k)$ be the number of distinct planar graphs having $k$
$\Phi$ lines (\cf\ \eq{pertexp}). The dressed gluon propagator in
Landau gauge is then (with standard gluon color indices):
\beq
\label{dressedgluon0}
i D^{\mu\nu}_{ab}(p) = \frac{-i}{p^2} \, P_T^{\mu\nu}(p) \,
d\left(- \frac{2 \mu^2}{p^2}\right) \delta_{ab} \ \ \ {\rm with} \ \ \
d(x) = \sum_{k=0}^\infty n(k)\, x^k  \,.
\eeq

In appendix B we show that $n(k) = C_k\, C_{k+1}$,
where $C_{k}$ is a Catalan number
(see \eq{catalan} and \eq{ndiag}).
Mathematica then gives\footnote{The expression \eq{polfact} for
    $d(x)$ may be verified by noting that $h(x)$ satisfies the
    differential
    equation  $x(1-x)h'' + (2-x) h' + h/4 =0$ with the condition
    $h(0)=1$.} for the `gluon polarization function' $d(x)$:
\beq
\label{polfact}
d(x) = \sum_{k=0}^\infty C_k\, C_{k+1}\, x^k =
   \frac{1-h(16x)}{2x}  \ \ \ {\rm with} \ \ \
h(x) = {_2F}_1\left(-\frac{1}{2}, \frac{1}{2}, 2, x \right) \,.
\eeq
The dressed gluon propagator is thus, in Landau gauge,
\beq
\label{dressedgluon}
i D^{\mu\nu}_{ab}(p) = \frac{i}{4 \mu^2} \, P_T^{\mu\nu}(p) \,
\left[1- {_2F}_1\left(-\frac{1}{2}, \frac{1}{2}, 2, -\frac{32
        \mu^2}{p^2} \right) \right] \delta_{ab}
\eeq

 From the integral representation of the hypergeometric
function $h(x)$,
\beq
h(x) = \frac{4}{\pi} \int_0^1 du \, \sqrt{1-u^2}\, \sqrt{1- x\,u^2}
\label{intrep}
\eeq
we see that $h(x)$ has a branch cut for
$x\geq 1$. This implies that the gluon propagator
\eq{dressedgluon} has a cut for
\beq
   - 32 \mu^2 \leq p^2 \leq 0 \, ,
\label{cut}
\eeq
which, surprisingly, lies in the spacelike $p^2 < 0$ region.

The polarization function $d(x)$ approaches unity in the
limit of high gluon virtuality, $x= -2\mu^2/p^2 \to 0$:
\beq
d(x\to 0) = 1 + 2x + 10 x^2 + \morder{x^3} \label{shortdist}
\eeq
Thus the dressing effects are power suppressed at short distances, as
expected. In the long distance limit $p^2 \to 0$ ($x \to \infty$)
the function $d(x)$ vanishes,
\beq
d(x\to +\infty) =  \frac{8}{3\pi}
\, \frac{1}{\sqrt{-x}} + \frac{1}{2x} + \morder{\frac{1}{x\sqrt{x}}}
\label{longdist}
\eeq
which turns the $p^2=0$ pole of the free propagator into a
$1/\sqrt{p^2}$
singularity. As for the quark, this implies that the gluon
propagator decays with time $t$ as $1/\sqrt{t}$, see \eq{Dasymp}
of Appendix A.

\begin{figure}[t]
\begin{center}
\epsfig{figure=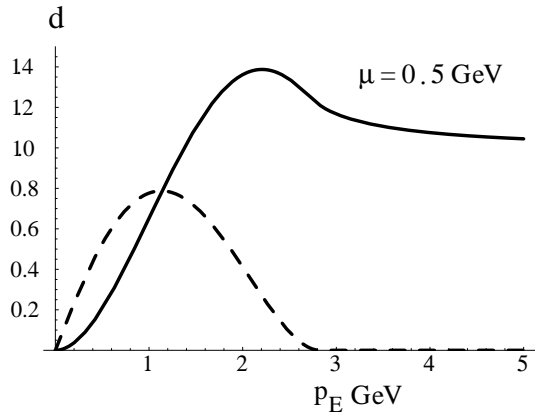,angle=0,width=7cm}
\end{center}
\caption{Real part (solid line) and
absolute value of the imaginary part (dashed line) of
   the gluon polarization function $d(x)$ in the
   euclidean region $p_E^2 \equiv -p^2 > 0$.}
\label{figgluon}
\end{figure}

We plot $d(-2\mu^2/p^2)$ for $p^2 < 0$ and $\mu = 500 \,{\rm MeV}$ in
Fig.~\ref{figgluon}.
The real part has a shape similar to the gluon propagator
found numerically in a Landau gauge lattice calculation, see 
Refs.~\cite{ds} or Fig.~3 of Ref.~\cite{Bonnet-gluon}. However, 
the presence in our expression
\eq{dressedgluon} of a branch cut in the euclidean region $p^2 < 0$
gives a negative imaginary part to the polarization function which
obscures the comparison.

Our result \eq{polfact} for the polarization function
$d(-2\mu^2/p^2)$ was
derived in Landau gauge ($\xi=0$ in \eq{gluonprop}). We now show that this
expression actually is independent of $\xi$, \ie, the dressed gluon
propagator in a general covariant gauge is
\beq
i D^{\mu\nu}_{ab}(p) = \frac{-i}{p^2} \, \left[ P_T^{\mu\nu}(p) \,
d \left(- \frac{2 \mu^2}{p^2}\right) + \xi
\frac{p^{\mu}p^{\nu}}{p^2}\right] \, \delta_{ab}
\label{covdressedgluon}
\eeq
The $\xi$-independence of $d(x)$ actually holds separately for any
diagram
contributing to $d(x)$, planar or non-planar.
This can be seen as follows, using the
standard single line color notation for the gluon and $\Phi$
propagators.
Let $k$ be the number of $\Phi$-lines of a diagram with given
topology. Since the free gluon propagator has a longitudinal part
$\propto \xi$,
\beq
i D_{0,ab}^{\mu\nu}(p) = \frac{-i}{p^2} \, \left(P_T^{\mu\nu}(p) + \xi
P_L^{\mu\nu}(p) \right)\, \delta_{ab}  \ \ ;\ \  P_L^{\mu\nu}(p)=
\frac{p^{\mu}p^{\nu}}{p^2}
\label{covfreegluon}
\eeq
the circled vertex \eq{vertex3} can contribute when $\xi \neq 0$.
There are
$2^{2 k}$ contributions for a given topology, since a chosen vertex
can be either circled or not.
When no vertex is circled the Lorentz structure brought by
the gluon line is
$\propto (P_T + \xi P_L)^{2 k +1} = P_T + \xi^{2 k +1} P_L$.
When at least one vertex is circled, consider the first vertex of this
type one meets by following
the gluon line (carrying momentum $p$) from left to right.
In all gluon propagators to the right of this vertex only the
longitudinal part $P_L$ contributes.
The $m$ gluon propagators to the left of this vertex give
$(P_T + \xi P_L)^m = P_T + \xi^m P_L$, \ie\ these $m$ propagators are
of the same type, either transverse or longitudinal.
Independently of the topological configuration,
$(P_T + \xi^m P_L)^{\mu \nu'}$ is always contracted with $p_{\nu'}$
($\nu'$ being the Lorentz index of the gluon line just before the
first circled vertex). As a result, in all $(2k +1)$ gluon
propagators, only the longitudinal part contributes when at least one
among the $2k$ vertices is circled. Since the circled vertex
has a negative sign relative to the uncircled one, the $\xi$-dependent
part brought by the $2^{2k}$ contributions is proportional to:
\beq
\xi^{2 k +1} P_L \sum_{n=0}^{2 k}  (-1)^n
{2k \choose n} = 0
\eeq
Only the $\xi$-independent (transverse) contribution remains,
which is obtained when no
vertex is circled. Thus the result found in Landau
gauge for $k \geq 1$ holds in all covariant gauges, establishing
\eq{covdressedgluon}.

\section{Photon self-energy}

In the previous section we saw that the dressed quark and gluon
propagators have no pole at $p^2=0$, only a $1/\sqrt{p^2}$ 
type singularity.
This implies a limited propagation length and should
soften the collinear $(p^2 \to 0)$ and infrared ($p \to 0$)
singularities of the perturbative expansion. In this section we study
the effects of dressing the quark loop contribution to the self-energy
of the photon.

\subsection{Coupling of zero-momentum gluons to a color singlet quark
loop \label{decoupsec}}

Spacelike photon fluctuations $\gamma(p) \to \qq \to \gamma(p)$
involve quark pairs of size $\sim 1/\sqrt{-p^2}$ (for massless
quarks). One might expect that such compact, short-lived color singlet
states would decouple from gluons of zero momentum, \ie, from the
$\Phi$ field we introduced in section~\ref{lagsec}. This charge
coherence effect is most easily seen for QED amplitudes. The standard
contribution of a bare electron loop to the photon self-energy is
\beq
\label{pi0}
\Pi^{\alpha\beta}(p) = ie^2 \int \frac{d^d k}{(2\pi)^d}
{\rm Tr} \left[\gamma^{\beta}S_e(k)\gamma^{\alpha} S_e(\bar k)\right]
\eeq
where $\bar k = k-p$. It is instructive to retain the mass in the
electron propagator
\beq
S_e(k) = \frac{1}{\Slash{k}-m}
\label{fermion}
\eeq
Now consider the attachment of an external zero-momentum photon with
Lorentz index $\mu_1$ to this loop. This contribution, denoted by
$\Pi_{\mu_1}^{\alpha\beta}(p)$, is given by two diagrams.
Using the identity
\beq
\label{identity}
- \frac{\partial}{\partial k^{\mu_1}} \left(
    \frac{1}{\Slash{k}-m} \right)
= \frac{1}{\Slash{k}-m} \, \gamma_{\mu_1} \,
\frac{1}{\Slash{k}-m}
\eeq
we note that $\Pi_{\mu_1}^{\alpha\beta}(p)$ is obtained
from \eq{pi0} by differentiating the integrand with respect to
$k^{\mu_1}$,
\beq
\label{pi1}
\Pi_{\mu_1}^{\alpha\beta}(p) = ie^2 \int \frac{d^d k}{(2\pi)^d} \,
\left[ -e \frac{\partial}{\partial k^{\mu_1}} \right] \,
{\rm Tr} \left[\gamma^{\beta}S_e(k)\gamma^{\alpha} S_e(\bar k)\right]
\eeq
Similarly, the QED amplitude for an electron loop with two external
zero-momentum photons is
\beq
\label{pi2}
\Pi_{\mu_1 \mu_2}^{\alpha\beta}(p) = ie^2 \int \frac{d^d k}{(2\pi)^d} \,
\left[ -e \frac{\partial}{\partial k^{\mu_1}} \right] \,
\left[ -e \frac{\partial}{\partial k^{\mu_2}} \right] \,
{\rm Tr} \left[\gamma^{\beta}S_e(k)\gamma^{\alpha} S_e(\bar k)\right]
\eeq

In Eqs.~\eq{pi1} and \eq{pi2} the integrand is given by a total
derivative. This generalizes to any number of
external zero-momentum photons. Thus the loop integral formally
vanishes, in agreement with intuition that zero-momentum photons do
not couple to a virtual $e^+e^-$ pair in a photon. The analogous proof
for QCD is somewhat more involved and is given in Appendix C.

However, this demonstration fails if the integral is singular and
thus ill-defined. As seen from \eq{identity} the insertion of a
zero-momentum photon increases the number of electron propagators
having the same momentum, which may cause infrared
singularities. Consider the photon self-energy contribution with $n$
zero-momentum photons attached to a (massive) electron loop. In this
amplitude there are diagrams with up to $n+1$ electron propagators of
the same momentum, \eg, $[1/(\Slash{k}-m)]^{n+1}$. After a Wick
rotation the integral has contributions at low $k$ of the form (for
even $n$)
\beq \label{irdiv}
A_n \sim \int d^4 k\, k^n \, \frac{k\cdot\bar k}{(k^2+m^2)^{n+1}}
\eeq
For a finite electron mass $m \neq 0$ the integrand is always regular
at $k = 0$ and, being a total derivative, the loop integral must
indeed vanish. On the other hand, for $m=0$ the integral is
apparently IR regular only for $n \leq 2$
(and thus vanishes for $n=2$ when the UV behaviour is dimensionally
regularized). For $n \geq 4$ the loop integral is IR singular and
hence ill-defined.

The physical reason for this infrared behaviour is that the coupling of
a photon to an $e^+e^-$ pair is proportional to the electric dipole
moment of the pair, which favors large dipole configurations in the
integrand. For $n \geq 4$ the integral becomes divergent at $k=0$.
The electron is then delocalized in space and
by the uncertainty principle the pair can have any size. 
At $k=p$ the positron is similarly delocalized.
This long-distance behaviour is a physical effect which cannot (or  
should
not) be removed, \eg, using dimensional regularization. In the case of
a finite electron mass the maximum size of the virtual pair is set by
$1/m$ and no infrared singularities occur.

This analysis also applies to QCD.
Zero-momentum gluons can couple to virtual $\qq$
states in the photon and the dressing affects the photon self-energy.
Using our dressed (chirally symmetric)
quark propagator \eq{csbcons}, and the dressed
$\gamma\qq$ vertex which we shall next derive, we find an infrared
regular loop integral. The dressing thus keeps the quark pair at
a finite separation $\sim 1/\mu$.  Singularities only appear if one
tries to Taylor expand in powers of $\mu^2$, implying that
non-analytic terms in $\mu^2$ appear in a $\mu \to 0$ expansion.

\subsection{Dressed quark-photon vertex}

\begin{figure}[t]
\begin{center}
\epsfig{figure=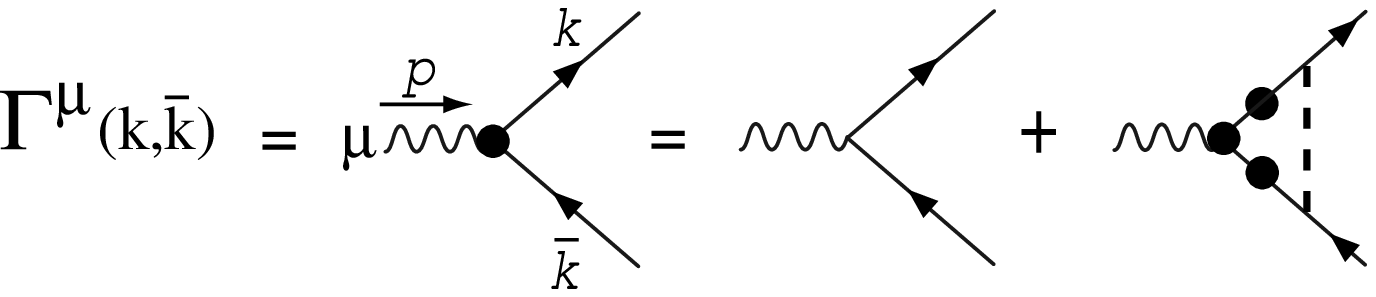,angle=0,width=11cm}
\end{center}
\caption{Implicit equation for the quark-photon vertex $\Gamma^{\mu}(k,
\bar k)$.}
\label{vertexDS}
\end{figure}

In our dressed tree approximation ($\ell =0$ in \eq{pertexp})
the $\gamma\qq$ vertex $\Gamma^{\mu}(k, \bar k)$ satisfies (at large
$N$) the implicit equation of Fig.~\ref{vertexDS},
\beq
\label{Gammagmuimplicit}
\Gamma^{\mu}(k, \bar k) =  \gamma^{\mu} - \halft\mu^2
\gamma^{\rho} S(k) \Gamma^{\mu}(k, \bar k) S(\bar k) \gamma_{\rho}
\eeq
where ${\bar k} = k-p$ and $p$ is the photon momentum. When the vertex
is expanded on its independent Dirac components \eq{Gammagmuimplicit}
reduces to a set of linear equations for the
components (\cf\ Eqs.~\eq{generalform} and \eq{lineqs}), with a unique
solution when the quark propagator $S$ is given. The explicit
expression \eq{Gammagmuexplicit0} for $\Gamma^{\mu}(k, \bar k)$ is
derived in Appendix D using the
chiral symmetry conserving quark propagator \eq{csbcons}.

Multiplying \eq{Gammagmuimplicit} by $p_\mu = (k-\bar k)_\mu$ and
rewriting \eq{S0} as
\beq
\halft\mu^2\gamma^\mu S(k)\gamma_\mu = S(k)^{-1} - \Slash{k}
\eeq
it is readily verified that the solution of \eq{Gammagmuimplicit}
respects the Ward-Takahashi identity
\beq
p_\mu\Gamma^{\mu}(k, \bar k)  = S(k)^{-1}- S(\bar k)^{-1}
\label{WT}
\eeq

For highly virtual momenta $k^2 \to \infty$ (at fixed $p$), we have
$S(k) \to 1/\Slash{k}$, $S(\bar k) \to 1/\Slash{\bar k}$
and \eq{Gammagmuimplicit} thus implies
\beq \label{asp2g}
k^2 \to \infty \ \ \Rightarrow \ \
\Gamma^{\mu}(k, \bar k) =
\gamma^\mu +\morder{\frac{\mu^2}{k^2}}
\eeq
This may also be verified from the explicit expression
\eq{Gammagmuexplicit0} of the dressed vertex.

\subsection{Dressed quark loop}

The dressed quark loop correction to the photon propagator is given
by the dressed quark propagator and quark-photon vertex as indicated
in Fig.~\ref{photonself}. In terms of the solutions
of the DS equations for the quark propagator \eq{S0} and vertex
\eq{Gammagmuimplicit} the photon self-energy correction
$\Pi^{\mu\nu}(p)$ is
\beqa
\label{selfg}
\Pi^{\mu\nu}(p) &=& ie^2 N \int \frac{d^d k}{(2\pi)^d}
{\rm Tr} \left[\gamma^{\nu}S(k)\Gamma^{\mu}(k,\bar k)S(\bar k)\right] \\
\label{selfg2}
&=& \frac{2}{d-2} \frac{ie^2 N}{\mu^2}
\int \frac{d^d k}{(2\pi)^d}
\Big\{\tr [(\Gamma^{\mu}(k,\bar k) - \gamma^{\mu})
\gamma^{\nu}] \Big\}
\eeqa
where we used the relation \eq{Gammagmuimplicit} for the vertex and
$\gamma^{\nu} = \gamma_{\rho}\gamma^{\nu}\gamma^{\rho}/(2-d)$
in $d$ dimensions.

\EPSFIGURE[h]{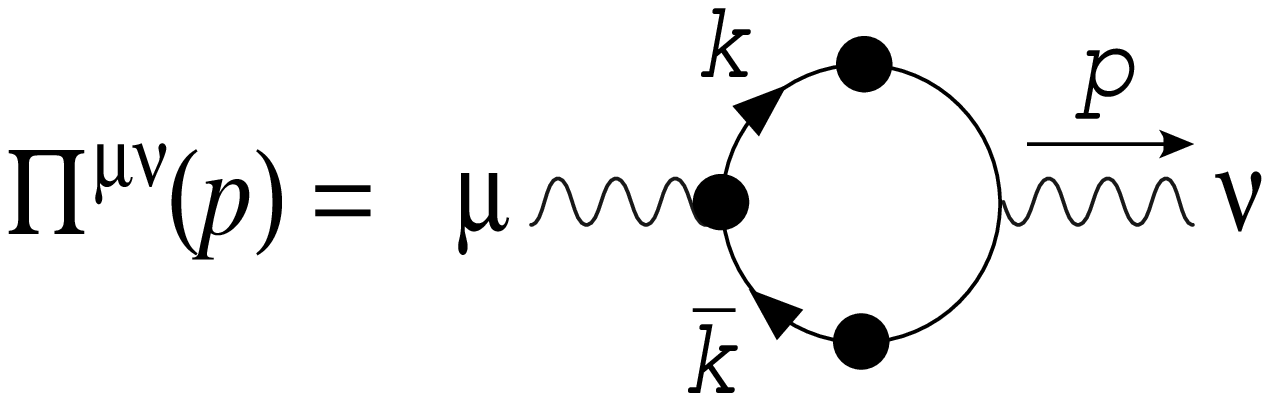,width=0.4\columnwidth}{Photon self-energy
$\Pi^{\mu\nu}(p)$ dressed by the $\Phi$ field.
The dressed propagators and vertex are indicated by a solid circle.
\label{photonself}}

We dimensionally regularize the standard ultraviolet divergence
which appears at zeroth order in $\mu^2$. At \order{\mu^2} the quark
loop couples to two photons and two $\Phi$ fields 
and is UV finite, but is still ill-defined in $d=4$ because of the
superficial logarithmic divergence of the loop integral.
Due to \eq{pi2} it vanishes when dimensionally regularized.

As is well-known, within dimensional regularization the 
Ward-Takahashi identity \eq{WT} implies a transverse photon self-energy
\beq
\label{transpi}
\Pi^{\mu\nu}(p) = \Pi(p^2)\,(p^2 g^{\mu\nu} - p^{\mu}p^{\nu})
\eeq
and a resummed photon propagator
\beq \label{photprop}
D_\gamma^{\mu\nu}(p) = \frac{1}{p^2[1-\Pi(p^2)]}\left(-g^{\mu\nu} +
\frac{p^{\mu}p^{\nu}}{p^2}\right)
\eeq
where $p^2\, \Pi(p^2) \to 0$ for $p^2 \to 0$.



\subsection{The dressed photon self-energy in euclidean space}

Here we consider in more detail the expression for
$\Pi(p^2)$, specified by \eq{selfg2} and \eq{transpi}, using
the quark propagator $S_1(p)$ given in \eq{csbcons} and the
quark-photon vertex \eq{Gammagmuexplicit0}. To avoid UV divergent
terms in the loop integral we subtract the terms of \order{\mu^0}
(\ie, the standard PQCD expression) and \order{\mu^2} (which
vanishes in dimensional regularization):
\beqa
\label{Pi}
{\widehat \Pi}(p^2) &\equiv& \Pi(p^2)  - \left. \Pi(p^2)
\right|_{\mu^0}  - \left. \Pi(p^2)\right|_{\mu^2}
\nn \\ && \\
&=& \frac{8 ie^2 N}{3p^2} \int \frac{d^4k}{(2\pi)^4}
\left\{ \frac{\mu^2 k^2 {\bar k}^2 a_k^2 a_{\bar k}^2}{1 - \mu^4
k^2 {\bar k}^2
           a_k^2 a_{\bar k}^2} - \frac{k\cdot\bar k\,  a_k a_{\bar k} +
\mu^2 k^2 {\bar k}^2
          a_k^2 a_{\bar k}^2}{1 + 2 \mu^2 k\cdot\bar k
           a_k a_{\bar k} + \mu^4 k^2 {\bar k}^2 a_k^2 a_{\bar k}^2}
\right. \ \
\nonumber \\
&& \hskip 7cm + \left. \frac{k\cdot\bar k}{k^2 {\bar k}^2} + \mu^2
    \frac{k\cdot\bar k}{k^4 {\bar k}^4}\, p^2  \right\} \nn
\eeqa
where the function $a_p$ is given in
\eq{aexpr}.

${\widehat \Pi}(p^2)$ has a complicated  analytic structure due to the
denominators of the integrand and the
square roots in $a_k$ and $a_{\bar k}$. We shall here only address its
properties for
euclidean momenta $p^2 < 0$, {\em assuming} that the loop integral
can be Wick rotated.

The denominator of the second term in the integrand of \eq{Pi} can be
estimated using $|k\cdot\bar k| \leq |k|\,|\bar k|$,
\beq
1 - 2 \mu^2 k\cdot\bar k
           a_k a_{\bar k} + \mu^4 k^2 {\bar k}^2 a_k^2 a_{\bar k}^2
\ge \left(1-\mu^2|k|\,|\bar k|a_k a_{\bar k} \right)^2 \ge 0
\eeq
where we reversed the sign of all dot products since the
momenta are now understood to be euclidean. This denominator
can thus vanish only if
\beq \label{vanden}
\mu^4 k^2 {\bar k}^2 a_k^2 a_{\bar k}^2 = 1
\eeq
which is also the condition for the denominator of the first term in
\eq{Pi} to vanish. For euclidean $k^2>0$ the function $a_k$ is negative,
\beq \label{aeucl}
a_k = \frac{1}{2\mu^2}\left(1-\sqrt{1+\frac{4\mu^2}{k^2}}\right) < 0
\eeq
The identity
\beq
\mu^2 a_k^2 k^2 = 1 + k^2 a_k
\eeq
then shows that $\mu^2 a_k^2 k^2 < 1$ (with a similar relation for
$\bar k$). Consequently the condition
\eq{vanden} cannot be fulfilled for finite euclidean momenta, and
the denominators in the expression \eq{Pi} for $\widehat\Pi$
are positive definite.

It is remarkable that the loop integral in \eq{Pi} is IR convergent.
For $k \to 0$, we have $\bar k \to -p$
and $a_k \sim -1/\mu\sqrt{k^2}$. Thus the first two terms of the
integrand
approach a constant, while the last two are IR safe.
As we noted above in \eq{irdiv}, the coefficients of $\mu^n$ in the
integrand become progressively more IR singular as $n$ increases.
The sum over all $n$ implied by the dressing nevertheless gives
a finite result.

The \order{\mu^4} contribution to ${\widehat \Pi}(p^2)$ in \eq{Pi} is
\beqa \label{UVpihat0}
\left.{\widehat \Pi}(p^2)\right|_{\mu^4} &=&
   \frac{16e^2N}{3p^2}  \, \mu^4 \\
&\times& \int \frac{d^4k}{(2\pi)^4}
\frac{k\cdot{\bar k}}{k^6{\bar k}^6} \, \left[k^4 + {\bar k}^4 - 2
k\cdot{\bar k} (k^2 + {\bar k}^2) - k^2{\bar k}^2 +2 (k\cdot{\bar k})^2
\right]  \nn
\eeqa
The integral is UV convergent, but has superficial linear and 
logarithmic divergencies for $k\to 0$ (and similarly for 
$\bar k\to 0$). The linearly divergent terms are
odd in $k$ and thus do not contribute to the integral. The  
logarithmically
divergent terms give
\beq \label{mu4div}
\left.{\widehat \Pi}(p^2)\right|_{\mu^4,IR} = 2 \cdot  
\frac{16e^2N}{3p^2}
\, \mu^4 \int \frac{d^4k}{(2\pi)^4}
\frac{1}{k^4 p^2} \, \left[1-4 \frac{(k\cdot p)^2}{k^2 p^2} \right] =0
\eeq
which vanishes due to the angular integration in $d=4$. Hence
the \order{\mu^4} term \eq{UVpihat0} is actually finite, and gives the
leading
behaviour of ${\widehat \Pi}(p^2)$ in the limit $p^2 \to \infty$,
\beq
{\widehat \Pi}(p^2) \ \ \mathop{\sim}_{p^2 \to \infty} \ \
-\frac{7\alpha N}{3 \pi} \, \frac{\mu^4}{p^4}
\label{UVpihat}
\eeq

We recall that the defining expression \eq{Pi} for ${\widehat \Pi}(p^2)$
is IR regular. A logarithmic singularity in the \order{\mu^4}
expression \eq{UVpihat0}
would have signalled an asymptotic behaviour $\propto
(\mu^4/p^4)\log(\mu^4/p^4)$.
Since according to \eq{irdiv} the IR behaviour becomes more singular
with the power $n$ of $\mu$ it is likely that the next-to-leading term  
in
the $p^2 \to \infty$ limit vanishes more slowly than  
$\morder{\mu^6/p^6}$.

\section{Concluding remarks}

We have formulated a modified perturbative expansion of QCD, where
the standard diagrams are dressed by
a constant external gluon field $\Phi_\mu^a$ which
is gaussian distributed in magnitude. We derived
explicit expressions for the dressed massless
quark and gluon propagators, as well as for the photon-quark vertex, at
lowest order in the standard loop expansion. We discussed some general
properties of the quark loop contribution to the photon self-energy.

Our formulation is explicitly gauge and Poincar\'e invariant, and
at short distances introduces only power suppressed
corrections to the standard PQCD results. The dressed
propagators have a branch point singularity instead of a pole at
$p^2=0$. Hence quarks and gluons cannot appear as asymptotic states,
in accordance with intuition that colored objects do not propagate
to infinite distances
in a color field. We also found that the dressed quark loop
contribution to the photon self-energy is regular in the infrared
euclidean
domain. This contrasts with the IR sensitivity of bare loops
with four or more soft external gluons, and implies that the physical
size of the dressed $\qq$ pair is governed by the scale $1/\mu$ related
to the strength of the field $\Phi$.

We introduced the external field $\Phi$ as a means
to study the qualitative effects of a gluon condensate on quark 
and gluon propagation.
While the features mentioned above are
encouraging, much work remains to be done to ascertain whether
this method yields results which are in accordance with general
principles. Our dressed quark and gluon propagators have a novel analytic
structure. It is especially surprising that the dressed gluon
propagator has a cut
in the {\it spacelike} $p^2 < 0$ region,
and it will be crucial to check that this behaviour is not
in contradiction with causality requirements.
The analytic properties of Green functions for confined fields
are largely unknown, an issue which the present framework may
help to clarify.
It will also be important to identify the asymptotic
states (if any!) of our framework, and to check whether an analytic
and unitary $S$-matrix can be defined.

\acknowledgments
We thank S.~J.~Brodsky, J.~P.~Guillet, M.~J\"arvinen and E.~Pilon
for interesting discussions, and T.~J\"arvi for help with a numerical calculation.

\break
\centerline{APPENDIX}

\appendix
\section{Asymptotic behaviour of the dressed quark and gluon
    propagators}

We give in this appendix the asymptotic time behaviour of the
dressed chirally symmetric
quark propagator $S_{1}(p)$ \eq{csbcons} and of the gluon
propagator in Landau gauge \eq{dressedgluon}.

The Fourier transformed propagators are defined by
\beq
S(t, \pvec) = \int_{-\infty}^{\infty} \frac{dp_0}{2\pi} S(p)
e^{-itp_0}\ \ ;\ \
D^{\mu\nu}(t, \pvec) = \int_{-\infty}^{\infty} \frac{dp_0}{2\pi}
D^{\mu\nu}(p) e^{-itp_0}
\eeq
We find
\beqa
S(t, \vec{p}) && \mathop{\sim}_{|t| \to \infty} \ \
-\frac{1+i}{2\sqrt{\pi}} \,\ \left( |\pvec| \gamma^0 -
         \pvec\cdot\vec{\gamma} \right)
\frac{\exp(-i|t\pvec|)}{\sqrt{|t\pvec|\mu^2}}
\label{Sasymp} \\
D^{\mu}_{\mu}(t, \vec{p}) && \mathop{\sim}_{|t| \to \infty} \ \
\frac{2\sqrt{2}}{\pi \sqrt{\pi}} (i-1) \,
\frac{\exp(-i|t\pvec|)}{\sqrt{|t\pvec|\mu^2}}
\label{Dasymp}
\eeqa
Since the two calculations are similar we give only the
derivation of \eq{Sasymp} for the quark propagator.

The quark propagator \eq{csbcons} may be written as
\beq
\label{B1}
S_{1}(p) = \frac{\Slash{p}}{2 \mu^2}
\left[ 1 - \frac{p^2- 4
      \mu^2}{\sqrt{p^2+\ieps}\sqrt{p^2-4\mu^2+\ieps}}\right]
\eeq
where the $\ieps$ prescription arises from the usual Feynman
prescription of the free quark propagator
$\Slash{p}/(p^2+\ieps)$.

The Fourier transform gives
\beqa
S(t, \pvec) &=&
\frac{(i \gamma^0 \partial_t - \pvec\cdot\vec{\gamma} )}{2\mu^2}
\left[\delta(t) + (\pvec^2 +4 \mu^2 + \partial_t^2) J(t, \pvec^2, \mu^2)
\right] \label{St} \\
J(t, \pvec^2, \mu^2) &=&  \int_{-\infty}^{\infty} \frac{dp_0}{2\pi}
\frac{\exp(-itp_0)}{\sqrt{p_0^2-\pvec^2+\ieps}\sqrt{p_0^2-\pvec^2
-4\mu^2+
           \ieps}} \label{J}
\eeqa
The function
$J(t, \pvec^2, \mu^2)$ can be evaluated using Feynman
parametrization,
\beq
\label{feynpar}
\frac{1}{\sqrt{A+\ieps}\sqrt{B+\ieps}} = \frac{1}{\pi}
\int_0^1\frac{dx}{\sqrt{x(1-x)}} \, \frac{1}{(1-x)A+xB+\ieps}
\eeq
and doing the $p_0$-integral using Cauchy's theorem.
The result is
\beq
J(t, \pvec^2, \mu^2) = \frac{e^{-i|t\pvec|}}{2i\pi}
\int_0^1\frac{dx}{\sqrt{x(1-x)}}
\frac{1}{\sqrt{\pvec^2+4x\mu^2}} \,
\exp\left[\frac{-4i|t|x\mu^2}{\sqrt{\pvec^2+4x\mu^2}+|\pvec|} \right]
\label{J2}
\eeq
The behaviour of $J(t, \pvec^2, \mu^2)$ for $|t| \to \infty$ can be
inferred
by noticing that the integrand in \eq{J2} is peaked at $x \to 0$ in
this limit. With the change of variable $y=2 |t|\mu^2 x/|\pvec|$ we
obtain
\beq
J(t, \pvec^2, \mu^2) \ \ \mathop{\longrightarrow}_{|t| \to \infty} \ \
\frac{e^{-i|t\pvec|}}{i\pi\sqrt{8|t\pvec|\mu^2}}
\int_0^{\infty} \frac{dy}{\sqrt{y}} e^{-iy} =
        - \frac{1+i}{4\sqrt{\pi}}
        \frac{e^{-i|t\pvec|}}{\sqrt{|t\pvec|\mu^2}}
\label{Jfinal}
\eeq
where we used
\beq
\int_0^{\infty}\frac{dy}{\sqrt{y}} \cos(y) =
\int_0^{\infty}\frac{dy}{\sqrt{y}} \sin(y) = \sqrt{\frac{\pi}{2}}
\eeq
Using \eq{Jfinal} in \eq{St} gives the asymptotic time behaviour
\eq{Sasymp}.

\section{Counting planar graphs}

In this appendix we calculate the number $n(k)$ of planar graphs
with $k$ internal $\Phi$-lines. This number appears in the
expression \eq{dressedgluon0} for the gluon self-energy.
As we already mentioned in section~\ref{gluesec}, in a planar graph
both ends of a $\Phi$-line attach to the gluon line either from above
or from
below. Let there be $a$ loops above and $b$ loops below the gluon
line, so that
\beq
k = a+b
\eeq
The number $C_a$ of ways to combine the $2a$ vertices with $a$ planar
loops above the gluon is independent of $b$, and is the same as for
the quark propagator. If each loop gives the same weight $x$ to the
quark
propagator $S(x)$ (ignoring its Dirac structure) then
\beq\label{Cdef}
S(x)= \sum_{a=0}^\infty C_a \, x^a
\eeq
The function $S(x)$ is determined by the DS equation which generates
all planar diagrams for the quark propagator (\cf\ Fig.~\ref{quarkDS}),
namely
$S(x) = 1 + xS^2(x)$ with $S(0)=1$. This gives
\beq
S(x) = \frac{1-\sqrt{1-4x}}{2x}
\eeq
A Taylor expansion shows that $S(x)$ is the generating function for
Catalan numbers. Thus the numbers $C_a$ in \eq{Cdef} are
\beq
\label{catalan}
C_a = \frac{1}{a+1} {2a \choose a}
\eeq
For a given ordering of the $2a+2b$ vertices on the gluon line we have
then $C_a C_b$ ways of drawing planar loops. The
number of distinct orderings of the vertices is given by the number
of ways to choose $2a$ vertices (regardless to their order)
from a set of $2k$ vertices, \ie, by the binomial factor
${2k \choose 2a}$.
The total number
$n(k)$ of distinct planar diagrams with $k$ loops to be used in
\eq{dressedgluon0} is thus:
\beq\label{ndiag}
n(k) = \sum_{a=0}^k {2k \choose 2a}
C_a C_{k-a} =  C_k\, C_{k+1}
\eeq
where the last equality may be derived as follows.
Rearranging factors in \eq{ndiag},
\beq
\label{B6}
n(k) = C_k \sum_{a=0}^k \frac{1}{a+1} {k \choose a} {k+1 \choose a}
   = \frac{C_k}{k+2} \sum_{a=0}^k {k \choose a} {k+2 \choose k+1-a}
\eeq
Comparing the coefficients of $x^{k+1}$ in the equivalent
expressions
\beqa
(1+x)^k (1+x)^{k+2} &=& \sum_{n=0}^{2k+2} x^n \sum_{a=0}^k
{k \choose a} {k+2 \choose n-a}
\nn\\
&&\label{binsum}\\
(1+x)^{2k+2} &=& \sum_{n=0}^{2k+2} {2k+2 \choose n}
x^{n} \nn
\eeqa
we get
\beq
\sum_{a=0}^k {k \choose a} {k+2 \choose k+1-a}
= {2k+2 \choose k+1}
\eeq
Using this in \eq{B6} we obtain
\beq
n(k) = \frac{C_k}{k+2} {2k+2 \choose k+1}
= C_k\, C_{k+1}
\eeq

\section{Decoupling of zero-momentum gluons from color-singlet quark
loops}
\label{sec:C0}

In section \ref{decoupsec} we gave a formal proof that
any number of external zero-momentum photons decouple
from an electron loop contribution to the photon self-energy,
due to charge coherence.
Here we shall extend this proof to the decoupling
of zero-momentum gluons from a (color singlet) quark loop correction
to the photon propagator.

\EPSFIGURE[h]{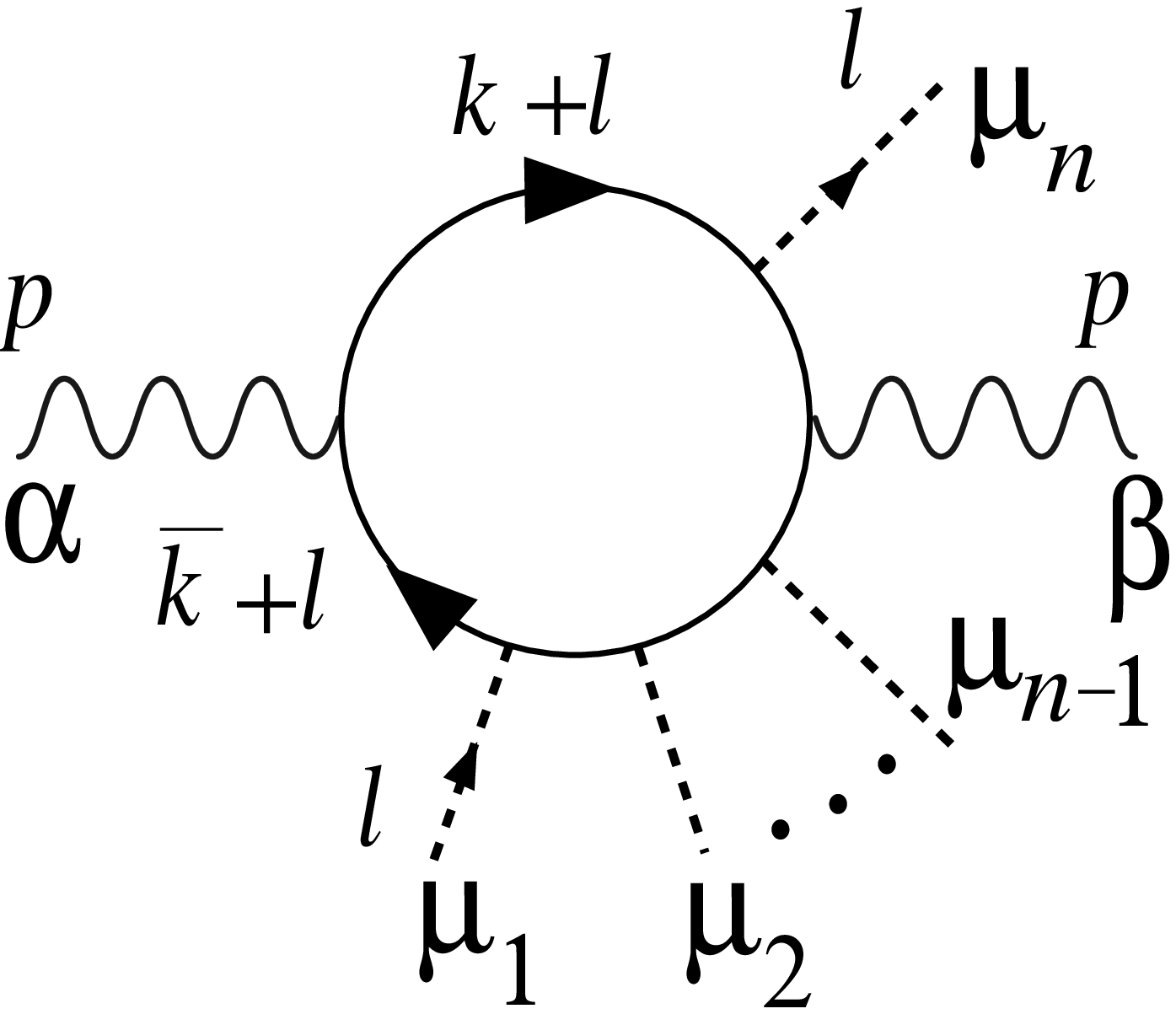,width=0.3\columnwidth}{A QED diagram
where the photons with vanishing four-momenta (denoted by dashed lines)
have a given cyclic ordering $\mu_1 \ldots \mu_n$. Lines $\mu_1$ and
$\mu_n$ are in the proof assigned momenta $l \neq 0$.\\
\label{formalproof}}

Let $\Pi_{\mu_1 \ldots \mu_n}^{\alpha\beta}(p)$ denote a quark loop
correction to a photon propagator with momentum $p$ and Lorentz
indices $\alpha, \beta$. The $n$ external, zero-momentum gluons
attached to the quark loop have Lorentz indices $\mu_1 \ldots \mu_n$,
and their color indices $a_1, \ldots, a_n$ are implicit.
The QED result
\beq
\label{decoupling}
\Pi_{\mu_1 \ldots \mu_n}^{\alpha\beta}(p) =0 \ \ {\rm for} \ \ n\geq 1
\eeq
also holds in QCD for
$n=1$ since ${\rm Tr} \, T^{a_1} = 0$, and for $n=2$ because the color
factor is then simply $\propto \delta^{a_1 a_2}$. For $n\geq 3$ the QCD
amplitude $\Pi_{\mu_1 \ldots \mu_n}^{\alpha\beta}(p)$ differs from the
QED one because the various diagrams are weighted by different color
factors of the type ${\rm Tr}\left[ T^{a_1}  \ldots T^{a_n}\right]$.
We now show that the part of the QCD amplitude which
corresponds to a {\it given} color factor formally vanishes by
itself. This ensures the vanishing of the complete QCD amplitude.

Consider the part of the corresponding QED amplitude which
is built from all diagrams that have the same cyclic
ordering of the external photon lines $\mu_1 \ldots \mu_n$,
where the indices are numbered by
following the fermion loop against the direction of the fermion
arrow. There are many diagrams of this type, which are
distinguished by the positions of the two photon lines with
momentum $p$ and indices $\alpha, \beta$
in the sequence of cyclically ordered external
photons (see Fig.~\ref{formalproof}).
We temporarily take the photons with indices $\mu_1$ and $\mu_n$ to
carry incoming momenta $l$ and $-l$, respectively, while
the photons $\mu_2 \ldots \mu_{n-1}$ have
vanishing four-momenta. Denoting the amplitude just described by
$T_{\mu_1 \ldots \mu_n}^{\alpha\beta}(p, l)$, we wish to
show that $T_{\mu_1 \ldots \mu_n}^{\alpha\beta}(p, 0) =0$.

The momentum $l$ may be taken to flow in the direction of the
fermion arrow between the vertices $\mu_1$ and $\mu_n$,
and not beyond. According to \eq{identity} the derivative
$-e \partial/\partial l^{\mu_{n+1}}$ applied
to $T_{\mu_1 \ldots \mu_n}^{\alpha\beta}(p, l)$ then generates all QED
diagrams with $(n+1)$ external zero-momentum photons
having the cyclic ordering $\mu_1 \ldots \mu_{n+1}$:
\beq
\left. -e \frac{\partial}{\partial l^{\mu_{n+1}}} T_{\mu_1 \ldots
    \mu_n}^{\alpha\beta}(p, l) \right|_{l=0} =
T_{\mu_1 \ldots  \mu_n \mu_{n+1}}^{\alpha\beta}(p, l=0)
\eeq
Assuming that the Taylor expansion
\beq
\label{taylor}
T_{\mu_1 \ldots \mu_n}^{\alpha\beta}(p, l) = T_{\mu_1 \ldots
    \mu_n}^{\alpha\beta}(p, 0) +  l^{\mu_{n+1}} \left.
\frac{\partial}{\partial
    l^{\mu_{n+1}}} T_{\mu_1 \ldots \mu_n}^{\alpha\beta}(p, l)
    \right|_{l=0} + \morder{l^2}
\eeq
is well-defined we obtain
\beq
\label{taylor2}
T_{\mu_1 \ldots \mu_n}^{\alpha\beta}(p, l)  =  T_{\mu_1 \ldots
    \mu_n}^{\alpha\beta}(p, 0) -\frac{1}{e} \, l^{\mu_{n+1}} T_{\mu_1
    \ldots  \mu_n \mu_{n+1}}^{\alpha\beta}(p, 0)  + \morder{l^2}
\eeq
It is readily seen from the charge conjugation symmetry
that
\beq
\label{furry}
T_{\mu_n \ldots \mu_1}^{\alpha\beta}(p, l)  = (-1)^n T_{\mu_1 \ldots
    \mu_n}^{\alpha\beta}(p, l)
\eeq
Reversing the order of $\mu_1 \ldots \mu_n$ in
\eq{taylor2} and recalling that the amplitudes were constructed to
be cyclically symmetric in the Lorentz indices of the zero-momentum
photons,
\beq
T_{\mu_n \ldots \mu_1 \mu_{n+1}}^{\alpha\beta}(p, 0)= T_{\mu_{n+1} \mu_n
    \ldots \mu_1}^{\alpha\beta}(p, 0)
\eeq
we get
\beq
\label{taylor3}
(-1)^n T_{\mu_1 \ldots \mu_n}^{\alpha\beta}(p, l)  =  (-1)^n T_{\mu_1
\ldots
    \mu_n}^{\alpha\beta}(p, 0) -\frac{1}{e} (-1)^{n+1} l^{\mu_{n+1}}
T_{\mu_1
    \ldots  \mu_n \mu_{n+1}}^{\alpha\beta}(p, 0)  + \morder{l^2}
\eeq
Comparing \eq{taylor2} and \eq{taylor3} we arrive at
\beq
T_{\mu_1 \ldots \mu_{n+1}}^{\alpha\beta}(p, 0)  = 0
\eeq
for $n\geq 2$. Since all QCD amplitudes with the same cyclic ordering
$\mu_1 \ldots \mu_n$ of the external gluons have the same color factor
${\rm Tr}\left[ T^{a_1}  \ldots T^{a_n}\right]$ this proof for QED
implies that \eq{decoupling} holds also in QCD.

\section{Quark-photon vertex $\Gamma^{\mu}(k, \bar k)$}
\label{sec:C}

In this appendix we solve the implicit equation \eq{Gammagmuimplicit}
for the $\gamma\qq$ vertex $\Gamma^{\mu}$ in the case of the
chirally invariant quark propagator \eq{csbcons},
\beq \label{ap}
S_{1}(p) = a_p\, \Slash{p}\ \ ;\ \ a_p =
\frac{1}{2\mu^2}\left(1-\sqrt{1-\frac{4\mu^2}{p^2}}\right)
\eeq
Using this expression for $S(p)$ \eq{Gammagmuimplicit} becomes
\beq
\label{C2}
\Gamma^{\mu}(k, \bar k) =  \gamma^{\mu} - \halft f
\gamma^{\nu} \Slash{k} \Gamma^{\mu}(k, \bar k) \Slash{\bar k}
\gamma_{\nu}
\eeq
where we introduced the dimensionful parameter $f$,
\beq
\label{f}
f = \mu^2 a_k a_{\bar k}
\eeq
Chiral and parity invariance restricts $\Gamma^{\mu}(k, \bar k)$ to
the form
\beq
\label{generalform}
\Gamma^{\mu}(k, \bar k) =  A_0 \gamma^{\mu} + A_1 k^{\mu} \Slash{k}
+ A_2 k^{\mu} \Slash{\bar k} + A_3 {\bar k}^{\mu} \Slash{k}
+ A_4 {\bar k}^{\mu} \Slash{\bar k} + i A_5 \gamma_5
\epsilon^{\mu}(\gamma, k, \bar k)
\eeq
where we defined
\beq
\epsilon^{\mu}(\gamma, k, \bar k) = \epsilon^{\mu\nu\rho\sigma}
\gamma_{\nu} k_{\rho} {\bar k}_{\sigma}
\eeq
We find the coefficients $A_i$ by inserting \eq{generalform} into
\eq{C2} and using
\beqa
\Slash{\bar k} \gamma^{\mu} \Slash{k} &=& k^{\mu} \Slash{\bar k} +
{\bar k}^{\mu} \Slash{k}  - k\cdot{\bar k} \, \gamma^{\mu} -  i \gamma_5
\epsilon^{\mu}(\gamma, k, \bar k)  \nonumber \\
i \gamma_5 \Slash{\bar k} \epsilon^{\mu}(\gamma, k, \bar k) \Slash{k}
&=&
- i \gamma_5 \epsilon^{\mu}(\gamma, k, \bar k)\, k\cdot{\bar k}
+ \gamma^{\mu} \left[k^2 {\bar k}^2 - (k\cdot{\bar k})^2 \right]
\nonumber \\ &&- k^{\mu} \left[ {\bar k}^2 \Slash{k} - k\cdot{\bar
           k}\, \Slash{\bar k}\right]
- {\bar k}^{\mu} \left[k^2 \Slash{{\bar k}} - k\cdot{\bar k} \,
         \Slash{k}\right]
\eeqa
This gives the conditions
\beqa \label{lineqs}
A_0 &=& 1 - f k\cdot{\bar k} \, (A_0 +k\cdot{\bar k} \, A_5)
+ f k^2{\bar k}^2 A_5 \nonumber \\
A_1 &=& f {\bar k}^2 (A_2 - A_5) \nonumber \\
A_2 &=& f (A_0 + k\cdot{\bar k}\, A_5 + k^2 A_1) \nonumber \\
A_3 &=& f (A_0 + k\cdot{\bar k}\, A_5 + {\bar k}^2 A_4) \nonumber \\
A_4 &=& f k^2 (A_3 - A_5) \nonumber \\
A_5 &=& - f (A_0 +k\cdot{\bar k} \, A_5)
\eeqa
with solutions
\beqa
A_0 = \frac{1 + f k\cdot{\bar k}}{1+2f k\cdot{\bar k}+f^2k^2{\bar
           k}^2} \ \ &;& \ \ A_5 = \frac{-f}{1+2f k\cdot{\bar
k}+f^2k^2{\bar
           k}^2} \nonumber \\
k^2 A_1 = {\bar k}^2 A_4 = - \frac{2fk^2{\bar k}^2}{1-f^2k^2{\bar
           k}^2} A_5 \ \ &;& \ \ A_2 = A_3 = -\frac{1+f^2k^2{\bar
k}^2}{1-f^2k^2{\bar
           k}^2} A_5
\eeqa
The result for $\Gamma^{\mu}(k, \bar k)$ then follows from
\eq{generalform}:
\beqa
\Gamma^{\mu}(k, \bar k)&=&
\frac{1}{1+2f k\cdot{\bar k}+f^2k^2{\bar k}^2} \left\{
(1 + f k\cdot{\bar k}) \gamma^{\mu} - f i\gamma_5 \,
\epsilon^{\mu\nu\rho\sigma} \gamma_{\nu} k_{\rho} {\bar k}_{\sigma}
\phantom{\frac{f^2}{f^2}} \right. \nonumber \\
&+& \left. \frac{2f^2}{1-f^2k^2{\bar k}^2}
(k^{\mu} \Slash{k} {\bar k}^2 + {\bar k}^{\mu} \Slash{\bar k} k^2)
+\frac{f(1+f^2k^2{\bar k}^2)}{1-f^2k^2{\bar k}^2}
(k^{\mu} \Slash{\bar k} + {\bar k}^{\mu} \Slash{k}) \right\}
\label{Gammagmuexplicit0}
\eeqa

Let us check that this expression for the vertex
satisfies the Ward-Takahashi relation \eq{WT}. Straightforward
algebra yields:
\beq \label{WT1}
p_{\mu}\Gamma^{\mu}(k, \bar k) = \Slash{k}
\frac{1-f{\bar k}^2}{1-f^2k^2{\bar k}^2} - \Slash{\bar k}
\frac{1-f k^2}{1-f^2k^2{\bar k}^2}
\eeq
Using \eq{f} and
Eq.~\eq{condS} (for $b=0$),
\beq \label{quad}
a_p - \frac{1}{p^2} = \mu^2 a_p^2
\eeq
we get
\beq
\frac{1-f{\bar k}^2}{1-f^2k^2{\bar k}^2} = \frac{1}{k^2 a_k} \ \ ;\ \
\frac{1-f{k}^2}{1-f^2k^2{\bar k}^2} = \frac{1}{\bar k^2 a_{\bar k}}
\eeq
Substituting these expressions in \eq{WT1} gives the Ward-Takahashi
relation \eq{WT}.

\end{document}